%% file: noise_v04c.tex
\documentclass[aps,preprint,groupedaddress,longbibliography]{revtex4-1}

\usepackage{graphicx,epsfig,amsmath,amssymb,color}

\begin{document}
\newcommand{\kT}[0]{k_{\textrm{B}}T}
\newcommand{\myvec}[1]{\boldsymbol{#1}}
\newcommand{\myten}[1]{\mathsf{#1}}

\title{The Power Spectrum of Ionic Nanopore Currents: The Role of Ion Correlations}


\author{Mira Zorkot}
\author{Ramin Golestanian}
\email{ramin.golestanian@physics.ox.ac.uk}
\author{Douwe Jan Bonthuis}
\email{douwe.bonthuis@physics.ox.ac.uk}
\affiliation{Rudolf Peierls Centre for Theoretical Physics, Oxford University, Oxford, OX13NP, United Kingdom}


\date{\today}

\begin{abstract}
We calculate the power spectrum of electric field-driven ion transport through cylindrical nanometer-scale pores using both linearized mean-field theory and Langevin dynamics simulations. 
With the atom-sized cutoff radius as the only fitting parameter, the linearized mean-field theory accurately captures the dependence of the simulated power spectral density on the pore radius and the applied electric field.
Remarkably, the linearized mean-field theory predicts a plateau in the power spectral density at low frequency $\omega$, which is confirmed by the Langevin dynamics simulations at low ion concentration.
At high ion concentration, however, the power spectral density follows a power law that is reminiscent of the $1/\omega^{\alpha}$ dependence found experimentally at low frequency.
Based on simulations with and without ion-ion interactions, we attribute the low-frequency power law dependence to ion-ion correlations.
Finally, we show that the surface charge density has no effect on the frequency dependence of the power spectrum.
\end{abstract}

\maketitle

\section{Introduction}

Measuring the ionic current passing through a nanometer-scale membrane pore has emerged over the past couple of decades as a versatile technique to study nanometer-scale transport processes \cite{2007_Dekker}.
In particular, nanopores are being used to study the properties of translocating biological molecules; to count them, measure their size and translocation velocity, 
or even -- in the case of DNA --  determine their sequence \cite{1996_Kasianowicz, 2000_Deamer, 2000_Meller}.
The sensitivity of such measurements is limited by the noise level, which typically increases with decreasing frequency $\omega = 2 \pi f$ \cite{1988_Weissman_RevModPhys}.
A high signal-to-noise ratio, as well as a consistent noise spectrum across individual devices, are among the requirements for a nanopore system to be of practical use.
Apart from limiting the measurement accuracy, however, the noise level can sometimes be used to study the microscopic properties of a nanofluidic system, such as the adsorption of molecules on the walls of a nanometer-scale cavity
\cite{2011_Singh-Lemay}.
At low frequency, the power spectrum $S\left(\omega\right)$ of the ionic current through an electrolyte-filled nanopore typically follows a power law $S\left(\omega\right) \propto 1/\omega^{\alpha}$ with $\alpha \approx 1$, 
which is referred to as pink noise, or flicker noise, or $1/f$ noise \cite{2015_Heerema, 2008_Smeets_PNAS}.
At high frequency, $S\left(\omega\right)$ is dominated by thermal or capacitive noise \cite{2008_Smeets_PNAS, 2007_Tabard-Cossa}.

Ever since the discovery of the near-universal appearance of pink noise in electrical systems, its microscopic origins have been heavily debated.
Measurements in protein channels indicate that the pink noise in biological nanopores is related to small conformational fluctuations of the open nanopore \cite{2000_Bezrukov_PRL, 1997_Wohnsland}.
These results are supported by experiments on nanopores in membranes with varying flexibility, showing pink noise in pores in flexible membranes, and a quadratic noise spectrum ($\alpha = 2$)
in pores in solid membranes \cite{2002_Siwy-Fulinski_PRL}.
Other explanations of pink noise include surface charge fluctuations \cite{2009_Hoogerheide} and the rectifying nature of conical pores \cite{2009_Powell-Siwy_PRL}.
Alternatively, the noise is thought to originate not in the properties of the nanopore, but instead in the nonequilibrium dynamics of charge carriers under confinement.
The first of these possible explanation stems from Hooge, whose measurements on homogeneous metal samples showed that the amplitude of the pink noise is proportional to the concentration of mobile charge carriers \cite{1970_Hooge}.
Further evidence is provided by the observation that stationary confining walls affect the velocity autocorrelation function of colloidal particles at long time scales \cite{1997_Hagen}.
Fractional Brownian motion, which can be used to describe pink noise,
also governs subdiffusive motion of molecules under confinement \cite{2010_Jeon-Metzler}.
Based on careful analysis of noise measurements in solid-state nanopores filled with ionic liquids, 
Tasserit \textit{et al.} reached the conclusion that pink noise is caused by a cooperative effect in the ionic motion \cite{2010_Tasserit}.
In computer simulations, pink noise has only been observed in nanochannels when the single-file motion is enforced artificially \cite{2008_Fulinski}.

Identifying the source of noise in nanopores is essential for efforts to optimize experimental setups, and to design nanopore systems for use in large-scale technological applications.
Moreover, a comprehensive understanding of the power spectrum of nanopore ion currents would constitute the basis of a new probe of the pore's microscopic properties,
allowing researchers to extract a wealth of information from a part of the measured signal that has traditionally been discarded.
However, a systematic theoretical investigation of the noise spectrum in nanopores filled with an aqueous electrolyte has been lacking so far.

In this paper, we present the first comprehensive theoretical analysis of the nonequilibrium noise spectrum of an ion current through a rigid nanopore
using implicit water.
We derive an analytical expression for the noise spectrum in the mean-field regime, providing a tool to analyze and interpret experimental results in both the high and the low frequency limits.
We compare our analytical expressions with the results of Langevin dynamics simulations, showing excellent agreement at high frequency.
Although the low frequency results follow our theoretical linearized mean-field prediction at low ion concentration, 
at high ion concentration the simulation results deviate from the linearized mean-field prediction, and show a nonzero $\alpha$, indicative of pink noise.
Our results suggest that ion correlations are a source of pink noise in nanopores.

\section{Linearized Mean Field Theory}

\begin{figure}
\includegraphics[width=0.5\columnwidth]{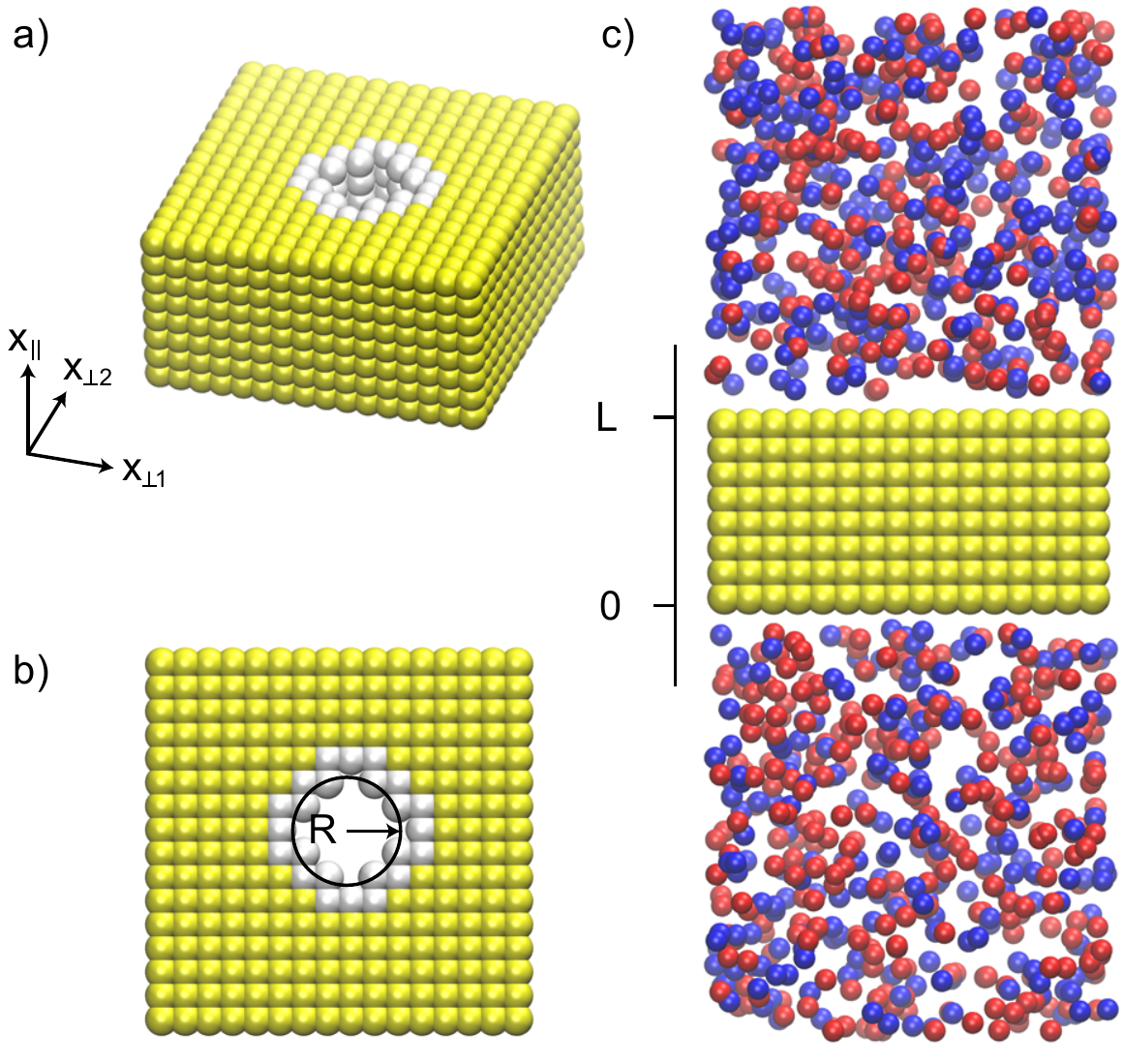}
\caption{
Snapshots from our simulations with (a) the membrane with the nanopore,
(b) the top view of the membrane with the radius $R$ indicated, 
and (c) the side view of the simulation box with the membrane thickness $L$ indicated.
The red and blue particles depict the charged ions, 
the yellow particles depict the membrane, 
and the white particles depict the pore wall that might be charged or uncharged.
In (a) and (b) the ions are not shown.
}
\label{fig:setup}
\end{figure}

We consider a system consisting of a cylindrical nanopore of length $L$ and radius $R$ connecting two reservoirs.
Because the ion current through the system is ultimately determined by the flux through the pore, 
we base our model on the ion flux density $\myvec{J}^{\pm}\left(\myvec{x},t\right)$ inside the nanopore,
with $\myvec{x}$ denoting the position in three dimensions and $t$ denoting the time.
We model the electrolyte filling the nanopore as ions of valency $\pm 1$ in implicit water.
Note that the model implies that nonlinear terms originating in the coupling of the ionic motion to the fluid velocity are ignored.
The ion concentrations $C^{\pm}\left(\myvec{x},t\right)$ are governed by the continuity equation,
\begin{equation} \label{eqn:fick}
\frac{\partial C^{\pm}}{\partial t} + \nabla \cdot \myvec{J}^{\pm} = 0.
\end{equation}
The corresponding flux densities $\myvec{J}^{\pm}\left(\myvec{x},t\right)$ are given by the Nernst-Planck equation,
\begin{equation} \label{eqn:nernst-planck}
\myvec{J}^{\pm} = -D^{\pm}  \nabla C^{\pm} \mp D^{\pm}C^{\pm} \frac{e \myvec{E}}{k_{B}T} + \myvec{\eta},
\end{equation}
where $\myvec{E}\left(\myvec{x},t\right)$ is the applied electric field, $e$ denotes the ion charge, 
and $\myvec{\eta}\left(\myvec{x},t\right)$ denotes the thermal noise that accounts for fluctuations in the environment, most importantly the effect of collisions between water molecules and ions. 
From here on, we restrict ourselves to the case where the ion diffusion coefficients are equal, $D^{+} = D^{-} = D$, which is accurate for many common salts, such as KCl and KNO$_3$.
Because of the cylindrical geometry we have only two independent coordinates, one parallel to the length of the pore ($\parallel$) an one perpendicular to the pore wall ($\perp$).
We are interested in the current in parallel direction in response to an electric field that is constant inside the pore and nonzero only in parallel direction $\myvec{E}\left(\myvec{x},t\right) = E_{\parallel}$.
We determine the Fourier transforms (denoted by $\widetilde{...}$) of Eqs. \ref{eqn:fick} and \ref{eqn:nernst-planck} as a function of the wave vectors $q_{\parallel}$ and $q_{\perp}$ and the frequency $\omega$ \cite{supp}, and calculate the power spectral density $S\left(\omega\right)$ from the Fourier-transformed current density $\widetilde{J}_{\parallel}^{+}\left(q_{\parallel},q_{\perp},\omega\right) - \widetilde{J}_{\parallel}^{-}\left(q_{\parallel},q_{\perp},\omega\right)$.
In terms of the Cartesian coordinates (Fig. \ref{fig:setup}a), the perpendicular wave vector is defined by $q_{\perp}^2 = q_{\perp1}^2 + q_{\perp2}^2$.
Multiplying by the Fourier-transformed lateral surface area $\widetilde{A}(q_\perp)$ of the pore,
and integrating over all wave vectors inside the nanopore,
we arrive at the power spectral density $S\left(\omega\right)$ \cite{supp},
\begin{equation} \label{eqn:psd}
S = 2 \int^{\varLambda^{-1}}_{R^{-1}} \frac{\textrm{d}^2 q_{\perp}}{(2\pi)^2} 
      \int^{\varLambda^{-1}}_{L^{-1}} \frac{\textrm{d}   q_{\parallel}}{2\pi} \, \widetilde{A}^2 \, \langle |\widetilde{J}_{\parallel}^{+}-\widetilde{J}_{\parallel}^{-}|^2\rangle
\end{equation}
with $R$ and $L$ being the radius and the length of the pore, respectively, and $\Lambda$ being the cut-off length, which is of the order of the particle size.
The factor $2$ stems from the integral over negative wave vectors $q_{\parallel}$.
The ensemble average is denoted by $\langle...\rangle$ and
the squared area function $\widetilde{A}(q_\perp)^2$ is given by \cite{supp}
\begin{equation} \label{eqn:area}
\widetilde{A}^2 = \left(2\pi\right)^2 \left[\frac{R^2}{q_{\perp}^2} - \frac{2R\sin{R q_{\perp}}}{q_{\perp}^{3}} - \frac{2 \cos{R q_{\perp}}}{q_{\perp}^{4}} + \frac{2}{q_{\perp}^{4}}  \right] .
\end{equation}
The average modulus square Fourier-transformed current density $\langle|\widetilde{J}_{\parallel}^{+}\left(q_{\parallel},q_{\perp},\omega\right)-\widetilde{J}_{\parallel}^{-}\left(q_{\parallel},q_{\perp},\omega\right)|^2\rangle$ is calculated from Eqs. \ref{eqn:fick} and \ref{eqn:nernst-planck} as \cite{supp}
\begin{equation} \label{eqn:current}
\begin{split}
\langle| & \widetilde{J}_{\parallel}^{+} - \widetilde{J}_{\parallel}^{-}|^2\rangle = \\
& \frac{8 D C_{0}\Big[\frac{eE_{\parallel}}{kT}\Big]^2 \Big[\frac{\omega^2}{D^2} + q_{\perp}^{4}\Big] \Big[q_{\perp}^{2} + q_{\parallel}^{2}\Big]} 
       {\Big(\Big[q_{\perp}^{2} + q_{\parallel}^{2}\Big]^2  \!\! + \Big[\frac{ eE_{\parallel} }{kT}\Big]^2 \! q_{\parallel}^2 \! - \frac{\omega^2}{D^2} \! \Big)^2 
     \!\!  + 4 \frac{\omega^2}{D^2} \Big[q_{\perp}^{2} + q_{\parallel}^{2}\Big]^2},
\end{split}
\end{equation}
with $C_0$ being the bulk ion density.
Together with Eqs. \ref{eqn:area} and \ref{eqn:current}, Eq. \ref{eqn:psd} is solved numerically to get the linearized mean-field prediction.

Remarkably, Eq. \ref{eqn:current} -- and therefore $S\left(\omega\right)$ -- is independent of the frequency $\omega$ in the limit $\omega \to 0$,
in clear contradiction to the low-frequency $S\left(\omega \to 0\right) \propto 1/\omega^{\alpha}$ behavior observed experimentally.
It is interesting to note that the same calculation in one dimension leads to $S\left(\omega \to 0\right) \propto \omega^2$, 
which is incompatibly not only with experimental results, but also with effectively one-dimensional simulations \cite{2008_Fulinski}.
At vanishing electric field $E_{\parallel}$, the noise spectrum becomes independent of $\omega$.

\section{Simulations}

We perform Langevin dynamics simulations of ions passing a membrane pore under influence of an applied potential difference using the molecular simulation package Espresso \cite{2006_Limbach}.
The Langevin equation for particle $i$ reads
\begin{equation} \label{eqn:langevin}
m_{i} \frac{\partial \myvec{u}_{i}}{\partial t} = -\sum_{j\neq i} \nabla V_{ij} + \myvec{F}_{i} - {\gamma} \myvec{u}_{i} + \myvec{\xi}_{i},
\end{equation}
with $\myvec{u}_{i}\left(t\right)$ being the particle's velocity in units \AA/$\tau$, $\gamma = 1$~$k_{B} T \tau$/\AA$^2$ being the friction coefficient,
with $\tau$ being the time scale and $k_{B}T$ being the thermal energy.
The random force $\myvec{\xi}_{i}\left(t\right)$ satisfies $\langle \xi_{i}\left(t\right)\xi_{i}\left(t'\right)\rangle = 6 k_{B} T \gamma \delta\left(t-t'\right)$.
Because the only dynamical particles in our simulation are the ions, the particles are assumed to be of equal mass $m_{i} = 1$~$k_{B}T\tau^2/$\AA$^2$, effectively incorporating the mass into the time scale $\tau$.
The index $j$ runs over all other particles of the system and
the potential $V_{ij}\left(r_{ij}\right)$ (in units of $k_{B}T$) comprises the Coulomb interactions and the Lennard-Jones interactions, truncated and shifted to form a Weeks-Chandler-Andersen potential,
\begin{equation} \label{eqn:potential}
V_{ij}= l_{B}\frac{q_{i}q_{j}}{r_{ij}} + 4 \epsilon_{ij} \left[ \left(\frac{\sigma_{ij}}{r_{ij}}\right)^{12}-\left(\frac{\sigma_{ij}}{r_{ij}}\right)^{6}\right] + V_{\textrm{shift}},
\end{equation}
with $r_{ij}$ being the distance between particles $i$ and $j$ and $q_i$ being the charge of particle $i$ in units of the elementary charge $e$.
The Bjerrum length $l_{B} = e^2/\left(4\pi\varepsilon\varepsilon_0 k_B T\right)$, with $\varepsilon_0$ being the permittivity of vacuum, is set to $l_{B} = 7$~\AA, modeling electrostatic interactions in water with a dielectric constant of $\varepsilon = 80$.
The long-ranged electrostatic interactions are treated using P$^3$M (particle-particle-particle mesh) Ewald summation with an automatically determined real-space cutoff \cite{2006_Limbach}.
The Lennard-Jones potential is truncated and shifted to zero at $r_{ij} = 2^{1/6}\sigma_{ij}$, using $V_{\textrm{shift}} = \epsilon_{ij}/4$.
We use the Lennard-Jones parameters $\epsilon_{ij} = 0.5$~$k_{B}T$ for all particles, $\sigma_{ij} = 3$~\AA~for ion-ion interactions, and $\sigma_{ij} = 9$~\AA~for ion-membrane interactions.
The applied potential difference across the membrane is modeled by the external force $\myvec{F}_{i}$, where we use the approximation that the electric field exerts a force in $x_{\parallel}$-direction inside the pore only,
\begin{equation} \label{eq:external-force}
\myvec{F}_{i} = \Bigg\{ {\begin{array}{l l} 
q_{i} \myvec{E} & \quad \text{if $0<x_{\parallel}<L$ }\\
 0 & \quad \text{otherwise},
  \end{array}}
\end{equation}
with $\myvec{E}\left(\myvec{x},t\right) = E_{\parallel}$.
The electric field is varied between $E_{\parallel} = 0.3$~$k_{B}T/\left(e\text{\AA}\right)$ and $E_{\parallel} = 1.6$~$k_{B}T/\left(e\text{\AA}\right)$, which at $L=48$~\AA~corresponds to a potential difference between $0.4$~V and $2$~V.

The time scale $\tau$ is calibrated by calculating the conductivity in simulations of a bulk system with an applied homogeneous electric field, and comparing it to the conductivity of a solution of potassium chloride, leading to $\tau = 5$~ps \cite{supp}.

\begin{figure}
\includegraphics[width=0.5\columnwidth]{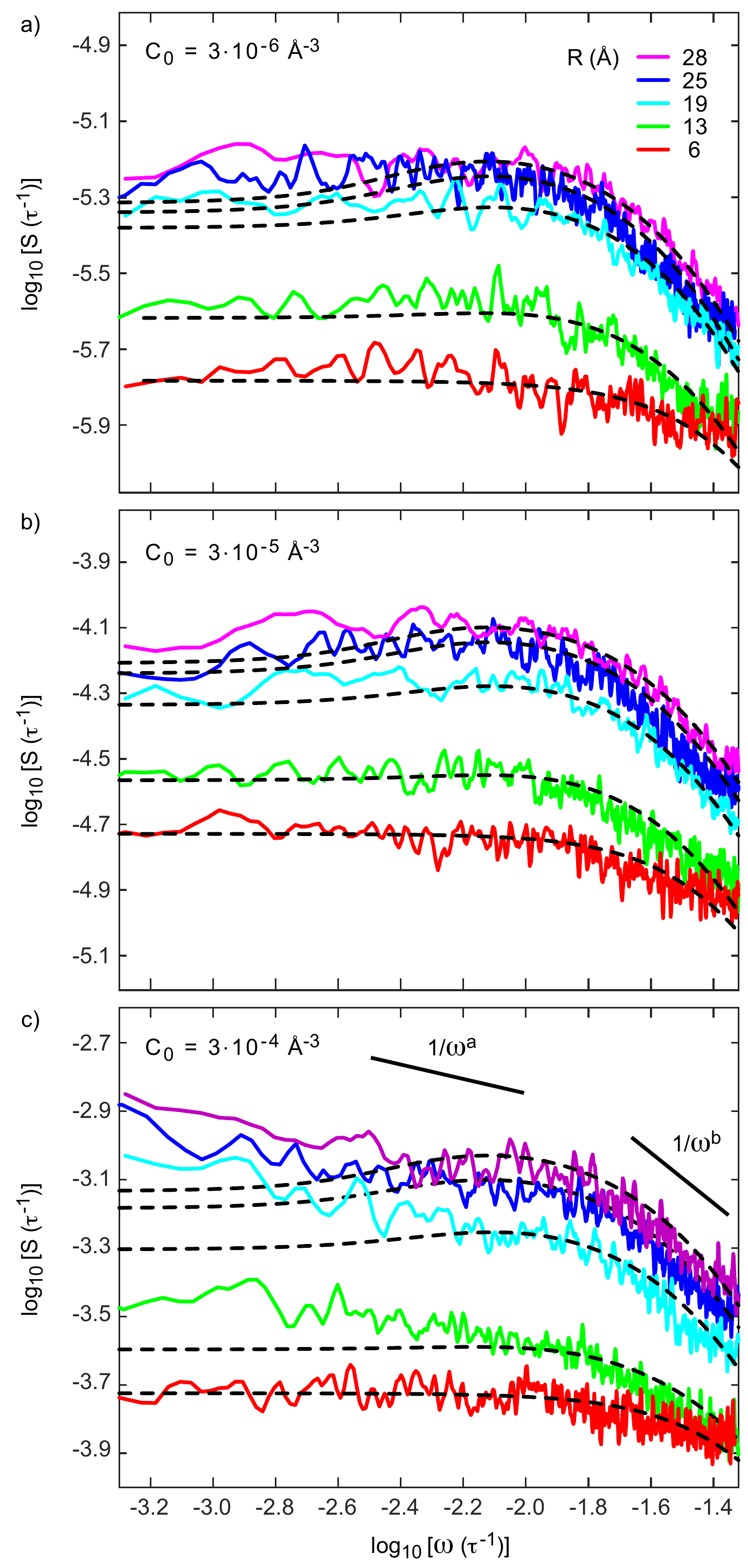}
\caption{
The power spectral density $S$ of the ionic current in units of the inverse time scale $\tau$, as a function of the frequency $\omega$ on a log-log scale, 
for ion concentrations (a) $C_0 = 3\cdot 10^{-6}$~\AA$^{-3}$, (b) $C_0 = 3\cdot 10^{-5}$~\AA$^{-3}$, and (c) $C_0 = 3\cdot 10^{-4}$~\AA$^{-3}$.
Solid curves denote $S$ from simulations for 5 different radii, and dashed lines denote fits with Eq. \ref{eqn:psd}.
The fitting parameter $\Lambda = 7$~\AA~for $R = 28$, $25$, and $19$~\AA, $\Lambda = 6$~\AA~for $R=13$~\AA, and $\Lambda=2.6$~\AA~for $R=6$~\AA for all concentrations.
The applied electric field is $E_{\parallel}=0.8$~\AA$^{-1}$.
}
\label{fig:concentrations}
\end{figure}

We construct a membrane of thickness $L=48$~\AA~consisting of membrane particles on a cubic lattice with a lattice constant of $6$~\AA, connected on both sides to a reservoir of width $W = 96$~\AA~and height $H = 100$~\AA.
Permeating the membrane, we construct a cylindrical pore of varying radii $R = \{6,13,19,25,28\}$~\AA.
The reservoirs and the pore are filled with positive and negative ions at three different concentrations, $C_0 = \{3\cdot 10^{-6},3\cdot 10^{-5},3\cdot 10^{-4}\}$~\AA$^{-3}$,
corresponding to $C_0 = \{0.005, 0.05, 0.5\}$~mol/l.
The membrane is verified to be impermeable to ions.
The membrane particles are frozen.
We use a simulation time step of $0.06 \tau$ and perform simulations of $3 \cdot 10^7$ time steps (Figs. \ref{fig:concentrations}-\ref{fig:electric-field}).
For Fig. \ref{fig:interactions} we use $1 \cdot 10^8$ time steps (with interactions) and $3 \cdot 10^8$ time steps (without interactions).
We discard $5000$ time steps for equilibration and calculate the current every $100$ time steps.
We use periodic boundary conditions in all directions.
We verify that the effect of the periodic boundary conditions, which affects fluctuations in general \cite{2014_Villamaina-Trizac}, is negligible in our case \cite{supp}.

We calculate the current $I_{\parallel} = \int (J^{+}_{\parallel}-J^{-}_{\parallel})\textrm{d}A$, 
equal to the integral of the current density over the lateral area $A$ of the pore,
from the integrated velocity in parallel direction
\begin{equation} \label{eqn:flux}
I_{\parallel} = \frac{1}{L} \sum_{i} 
\Bigg\{ {\begin{array}{l l} 
q_i\left(\myvec{u}_{i}\right)_{\parallel} & \quad \text{if $0 < x_{\parallel} < L$ }\\
 0 & \quad \text{otherwise},
  \end{array}}
\end{equation}
where the index $i$ runs over all positive and negative ions.

From the current we calculate the power spectrum using the Welch method with a Hamming window, an overlap of 0.5, and 100 windows \cite{1967_Welch}.

\section{Results and Discussion}

\textbf{Pore radius \& ion concentration.} --
In Fig. \ref{fig:concentrations} we show the power spectral density $S(\omega)$, calculated from the Langevin dynamics simulations, as a function of the frequency $\omega$ (solid lines).
The curves have been fitted with the linearized mean-field expression of Eq. \ref{eqn:psd} (dashed lines), using the high wave vector cutoff $\Lambda$ as the only fitting parameter.
Eq. \ref{eqn:psd} fits the curves with remarkable accuracy for all pore radii.
All curves show a transition around $\omega = 10^{-2}$~$\tau^{-1}$, and a power law decrease $S\propto 1/\omega^{b}$ at high frequency.
At even higher frequency, $S$ is dominated by white noise (not shown).
At low ion concentrations ($C_0 = 10^{-6}$~\AA$^{-3}$ and $C_0 = 10^{-5}$~\AA$^{-3}$, Figs. \ref{fig:concentrations}a-b), $S$ exhibits a plateau at low frequency,
as predicted by the linearized mean-field theory.
At higher ion concentration ($C_0 = 10^{-4}$~\AA$^{-3}$, Fig. \ref{fig:concentrations}c), the linearized mean-field theory still captures the high frequency simulation results.
At low frequency, however, we do not find a plateau, but $S$ continues to increase with decreasing frequency instead, following a power law $S\propto 1/\omega^{a}$.
This behavior is reminiscent of the $1/\omega^{\alpha}$ dependence of $S$ found experimentally at low frequency.

\begin{figure}
\includegraphics[width=0.5\columnwidth]{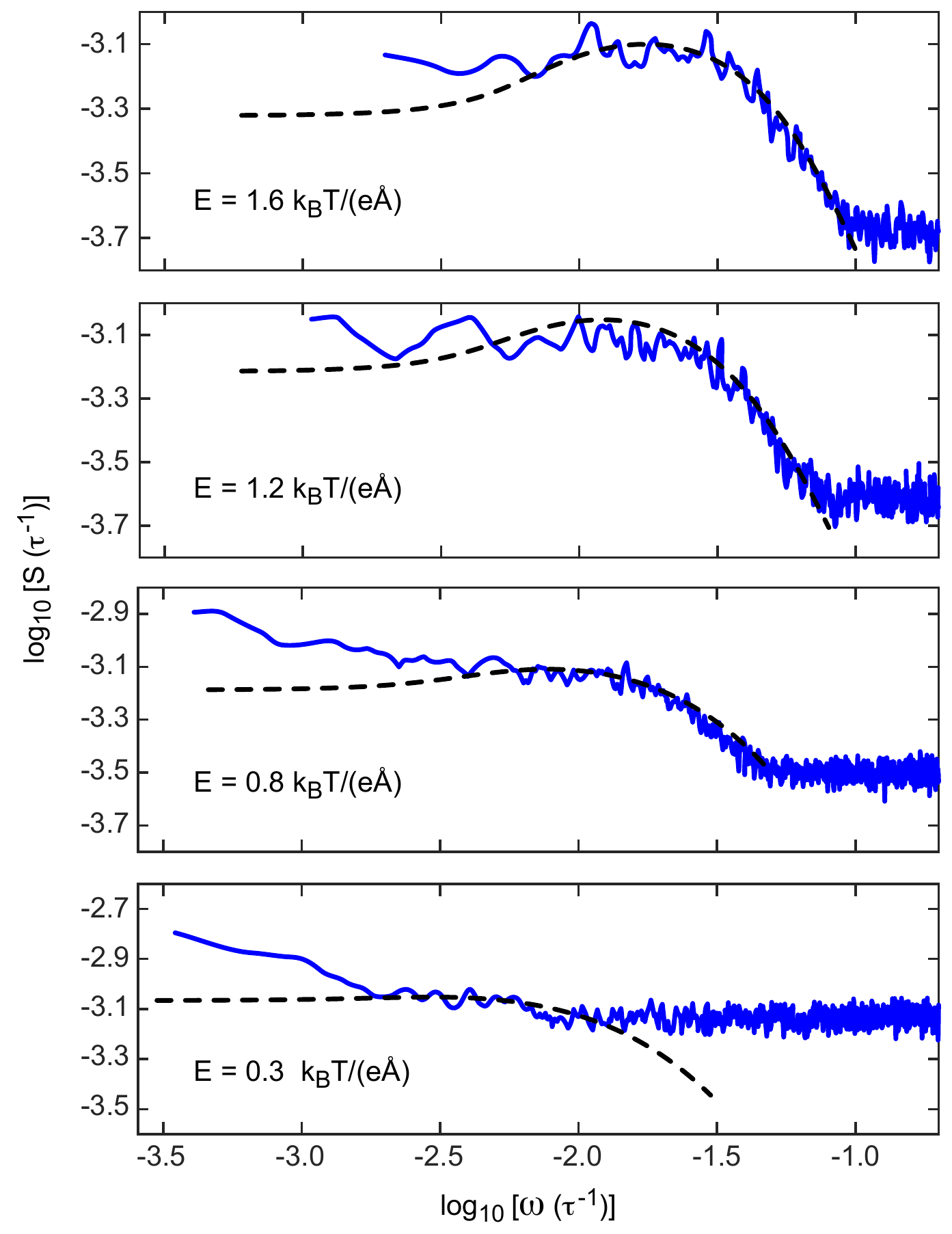}
\caption{
The power spectra of the ion current through a pore of $R = 25$~\AA~at a concentration of $C_0 = 3 \cdot 10^{-4}$~\AA$^{-3}$, at different applied electric field strengths inside the pore (0.3, 0.8, 1.2, and 1.6 $k_BT/\left(e\text{\AA}\right)$).
With increasing electric field strength, the transition frequency shifts upward, and the background white noise decreases.
}
\label{fig:electric-field}
\end{figure}

\textbf{Electric field.} --
We perform simulations of the pore of $R = 25$~\AA~at a concentration of $C_0 = 3\cdot 10^{-4}$~\AA$^{-3}$, varying the electric field from $0.3$ to $1.6$~$k_B T/(e\text{\AA})$ (Fig. \ref{fig:electric-field}).
Again, the linearized mean-field theory (dashed lines) fits the transition in the frequency domain very well for all electric fields, without further adjustable parameters.
The transition frequency shifts to higher frequencies for higher electric fields.
As predicted, the dependence of $S\left(\omega\right)$ on $\omega$ vanishes for vanishing electric field.

\begin{figure}
\includegraphics[width=0.5\columnwidth]{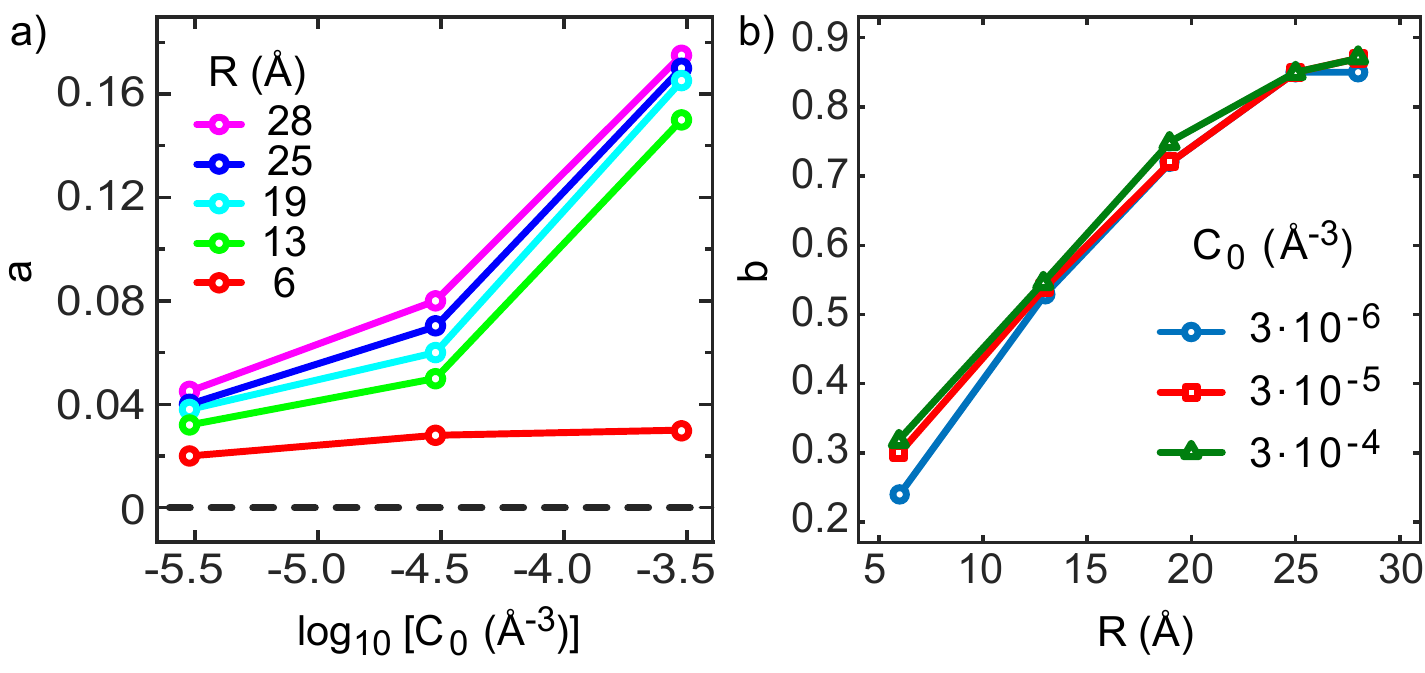}
\caption{
(a) The power $a$ of the fits $S\left(\omega\right) \propto \omega^{-a}$ ($\omega < 10^{-2}$~$\tau^{-1}$) as a function of concentration $C_0$.
The dashed line denotes the linearized mean-field prediction of Eqs. \ref{eqn:psd}--\ref{eqn:current}.
(b) The power $b$ of the fits $S\left(\omega\right) \propto \omega^{-b}$ ($\omega > 10^{-2}$~$\tau^{-1}$) as a function of the radius $R$.
The electric field is set to $E_{\parallel} = 0.8$~$k_B T/(e\text{\AA})$.
}
\label{fig:powers}
\end{figure}

\textbf{Limiting behavior.} --
We fit the low frequency regime ($\omega < 10^{-2}\tau^{-1}$) and the high frequency regime ($\omega > 10^{-2} \tau^{-1}$) with power laws with exponents $a$ and $b$, respectively (see Fig. \ref{fig:concentrations}c).
Apart from the case $R = 6$~\AA, the exponent $a$ shows a strong dependence on the ion concentration $C_0$ (Fig. \ref{fig:powers}a), increasing from $a < 0.05$ for $C_0 = 3\cdot 10^{-6}$ to $a > 0.15$ for $C_0 = 3\cdot 10^{-4}$.
For radii $R > 6$~\AA, $a$ is largely independent of $R$, suggesting that the power law behavior is an intrinsic property of nonequilibrium ion transport, rather than a consequence of the confinement.
In contrast, the exponent $b$ depends on the radius, but is independent of the ion concentration, as predicted by Eq. \ref{eqn:psd} (Fig. \ref{fig:powers}b).

\begin{figure}
\includegraphics[width=0.5\columnwidth]{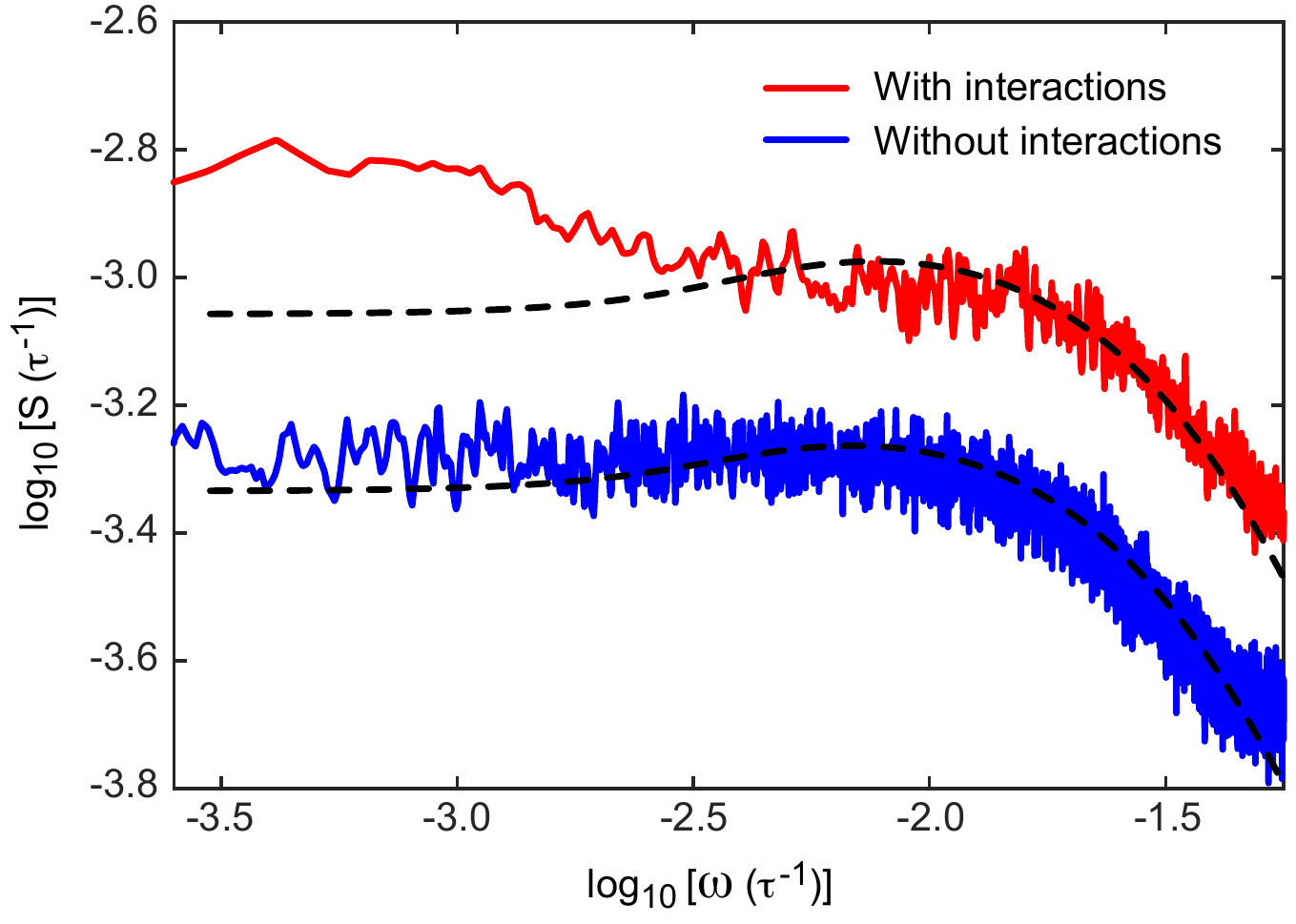}
\caption{
The power spectrum with the ion-ion interaction potential $V_{ion-ion}\left(r_{ij}\right)$ given by Eq. \ref{eqn:potential} (with interactions, red line) and with $V_{ion-ion}\left(r_{ij}\right) = 0$ (without interactions, blue line).
The spectrum is calculated in a $R = 25$~\AA~channel with $E_{\parallel} = 0.8$~$k_{B}T/(e\text{\AA})$.
}
\label{fig:interactions}
\end{figure}

\textbf{Ion correlations.} --
The strong dependence of $a$ on $C_0$, which is not predicted by mean-field theory, suggests a strong effect of ion correlations.
To investigate this low-frequency power law dependence of the power spectrum in more detail, we perform simulations with $V_{ion-ion} = 0$ for the ion-ion interactions at the high concentration $C_0 = 4 \cdot 10^{-4}$, eliminating ion correlations.
Ion-membrane interactions are left unchanged.
We perform extra long simulations in a pore with $R = 25$~\AA.
In Fig. \ref{fig:interactions}, we show the power spectrum obtained with $V_{ion-ion}\left(r_{ij}\right) = 0$, and compare it with the the power spectrum obtained with full interactions ($V_{ion-ion}\left(r_{ij}\right)$ given by Eq. \ref{eqn:potential}).
Clearly, with $V_{ion-ion}\left(r_{ij}\right) = 0$ the power spectrum saturates at low frequency, in agreement with the linearized mean-field prediction.
Therefore, the departure from linearized mean-field theory at high concentrations shown in Fig. \ref{fig:powers}a is due to the ion correlations caused by the ion-ion interaction term $V_{ion-ion}\left(r_{ij}\right)$.

\begin{figure}
\includegraphics[width=0.5\columnwidth]{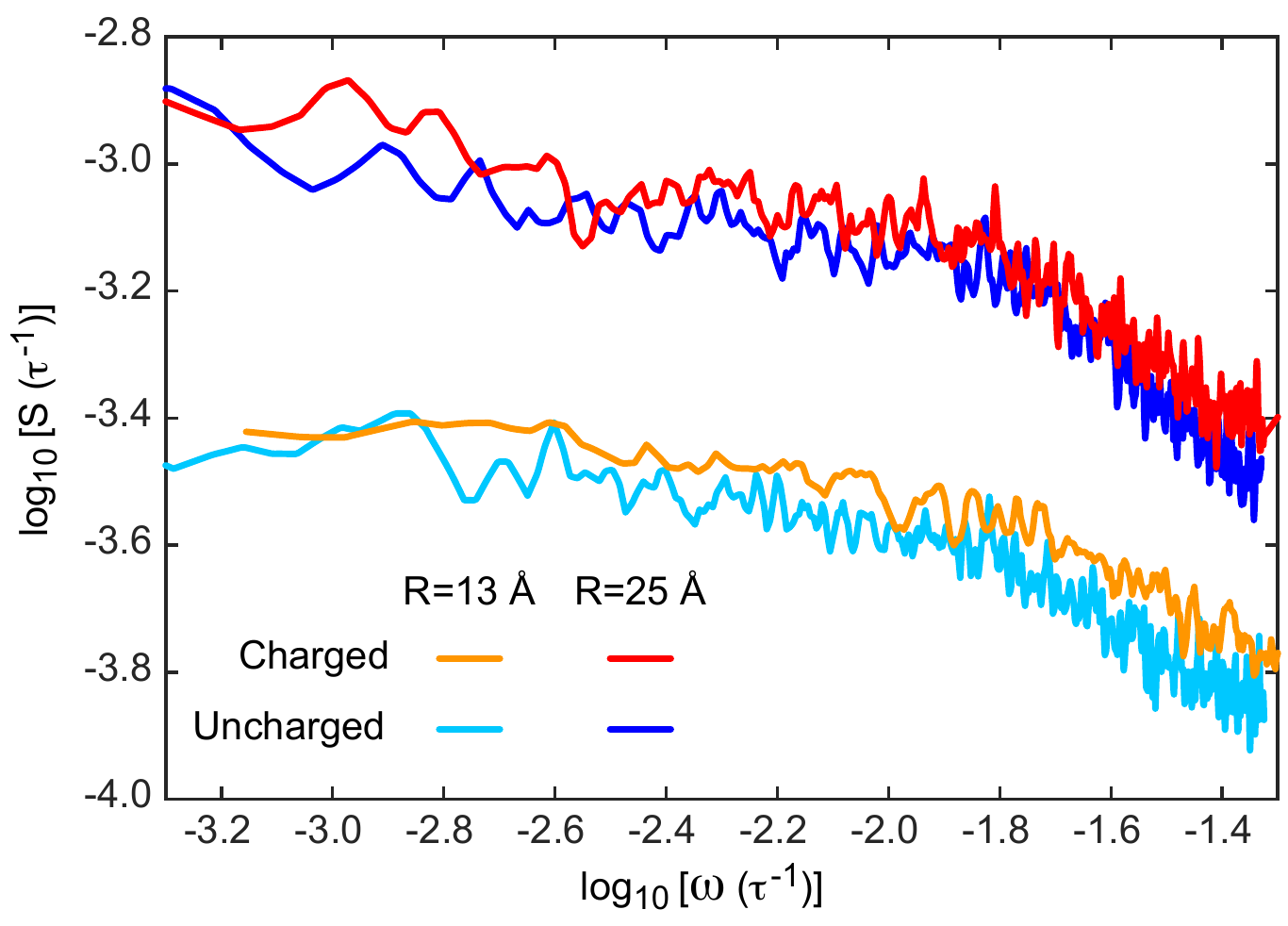}
\caption{The power spectral density of the ionic current through charged and uncharged pores of radii 25~\AA~and 13~\AA~at a concentration of $3\cdot 10^{-4}$~\AA$^{-3}$ and an electric field of $0.8$~$k_BT/(e\text{\AA})$.}
\label{fig:charge}
\end{figure}

\textbf{Surface charge.} --
Typically, both synthetic and biological aqueous pores have a finite surface charge density.
The interactions between the charged surface and the ions has been suggested as the source of the $1/\omega$ power law at low frequency \cite{2009_Powell-Siwy_PRL,2009_Hoogerheide}.
We investigate the effect of surface charges by allocating negative charges on the surface of the pores, balanced by extra positive ions in solution.
The charges are located at the membrane particles located in a cylindrical shell between $R < r < R + 6$~\AA, with the total charge equal to $Q = -19e$ for $R=13$~\AA~and $Q = -26e$ for $R=25$~\AA.
The corresponding power spectra are shown in Fig. \ref{fig:charge}.
No differences are observed in the shape, the transition frequency, or the power laws at low and high frequency, in agreement with Ref. \cite{2010_Tasserit}.
In the charged pores, the magnitude of the power spectral density is $20\%$ higher than in the uncharged pores, which we attribute to the increased ion density due to ions screening the surface charge.
The increase in the power spectral density corresponds indeed to the increased number of ions in the pore, which we determine independently.

\section{Conclusions}

We calculate the power spectrum of electric field driven ion transport through cylindrical nanometer scale pores using both linearized mean-field theory and Langevin dynamics simulations.
We derive that the linearized mean-field theory predicts a plateau in the power spectrum at low frequency, which is not found in experiments.
Furthermore, the linearized mean-field theory predicts a decreasing power law at high frequency, where the power depends on the applied electric field.
At low ion concentrations (0.005 mol/l - 0.05 mol/l), the Langevin dynamics simulations confirm the mean-field predictions with high accuracy, including the electric field dependence of the power law decrease at high frequency and the plateau at low frequency.
Our linearized mean-field expression fits the simulation data using only one adjustable parameter, the cut-off small length scale $\Lambda$, varying between 6~\AA~and 7~\AA~for all curves, apart from $R=6$~\AA, which gives $\Lambda = 3$~\AA.
At high ion concentration (0.5 mol/l), the simulated power spectrum is still accurately described by our linearized mean-field theory at high frequency, but at low frequency the simulation curves do not exhibit a plateau.
Instead, the power spectrum increases with decreasing frequency with a power law $S(\omega) \propto 1/\omega^a$, with $0.14 < a < 0.18$.
We attribute this deviation from the mean-field prediction to ion-ion correlations that are not present in the linearized mean-field theory.
Finally, we study the effect of a finite surface charge density on the inside of the nanopore. 
Contrary to reports in literature \cite{2009_Powell-Siwy_PRL,2009_Hoogerheide} we find no significant dependence of the power spectrum on the surface charge density.

\section{Appendix}

\textbf{Derivation of the governing equations.} --
To derive $\tilde{J}^+_{\parallel}\left(\myvec{q},\omega\right) - \tilde{J}^-_{\parallel}\left(\myvec{q},\omega\right)$,
we switch to index notation where $\alpha$, $\beta$, and $\gamma$ correspond to the three components of our coordinate system.
After applying a standard Fourier transform to Eqs. \ref{eqn:fick} and \ref{eqn:nernst-planck},
\begin{equation} \label{eq:er8}
\widetilde{J}^{z}_{\alpha} = \sum^{3}_{\beta=1} \left[ -\frac{iD}{\omega} \, q_{\alpha} q_{\beta} \widetilde{J}^z_{\beta} - \frac{D\,zeE_{\alpha}}{\omega\, k_BT} \, q_{\beta} \widetilde{J}^z_{\beta} - \widetilde{\eta}_{\alpha} \right],
\end{equation}
with $\widetilde{...}$ denoting the Fourier transform, $\myvec{q}$ being the wave vector, $\omega$ being the frequency, and $z = \pm 1$ denoting the ion charge. 
Rewriting Eq. \ref{eq:er8} leads to:
\begin{equation} \label{eq:er10}
\begin{split}
- \widetilde{\eta}_{\alpha} &=
\sum^{3}_{\beta=1} \widetilde{J}^z_{\beta} \left[\delta_{\alpha \beta} + \frac{iD}{\omega}\,q_{\alpha} q_{\beta} + \frac{D\,zeE_{\alpha}}{\omega \, k_BT} \, q_{\beta} \right] \\
&= \sum^{3}_{\beta=1} \widetilde{J}^z_{\beta} M_{\alpha\beta},
\end{split}
\end{equation}
where $M_{\alpha\beta}$ denotes the matrix
\begin{equation} \label{eq:er11}
M_{\alpha\beta}=\delta_{\alpha \beta} + \frac{iD}{\omega} \, q_{\alpha}q_{\beta} + \frac{D\,zeE_{\alpha}}{\omega \, k_BT} \, q_{\beta}.
\end{equation}
Combining Eqs. \ref{eq:er10} and \ref{eq:er11} and solving for $\widetilde{J}^z_{\gamma}$, we find
\begin{equation}\label{eq:er13}
\begin{split}
\widetilde{J}^z_{\gamma} &= \frac{1}{\rm{det}(M)}\left[\sum^{3}_{\beta=1} \left[-\frac{iD}{\omega}q_{\gamma}{q_{\beta}\widetilde{\eta}_{\beta}} - \frac{D\,zeE_{\gamma}}{\omega\,k_BT} \, q_{\beta} \widetilde{\eta}_{\beta} \right] \right. \\
 &\qquad + \left. \widetilde{\eta}_{\gamma}\sum^{3}_{\beta=1}{\left[\frac{1}{3} + \frac{iD}{\omega}q_{\beta}{q_{\beta}}
+ \frac{D\,ze{E}_{\beta}}{\omega \, k_BT} \, q_{\beta}\right]} \right].
\end{split}
\end{equation}
Because of our cylindrical symmetry, the flux has two unique components: Parallel ($\parallel$) and perpendicular($\perp$) (Fig. 1a). 
The electric-field is nonzero only in parallel direction $\myvec{E} = (0,0,E_{\parallel})$.
Because we are interested in the longitudinal flow of ions through the pore, we concentrate on the flux in the parallel direction:
\begin{equation} \label{eq:er14}
\begin{split}
\widetilde{J}^z_{\parallel}(\myvec{q},\omega) &=\frac{ \frac{iD}{\omega} \big[{q_{\perp 1}q_{\parallel}\widetilde{\eta}_{\perp 1}} + {q_{\perp 2}q_{\parallel} \widetilde{\eta}_{\perp 2}} + {q_{\perp 1}^{2} \widetilde{\eta}_{\parallel}} + {q_{\perp 2}^{2} \widetilde{\eta}_{\parallel}}\big] }{1 + \frac{iD}{\omega} \big[q_{\parallel}^2 + q_{\perp 1}^2 + q_{\perp 2}^2 \big] + \frac{D\,zeE_{\parallel}}{\omega\,k_BT} q_{\parallel}}\\
&\qquad + \frac{ -\frac{D\,zeE_{\parallel}}{\omega\,k_BT}\big[{q_{\perp 1} \widetilde{\eta}_{\perp 1}} + {q_{\perp 2} \widetilde{\eta}_{\perp 2}}\big] + \widetilde{\eta}_{\parallel}}{1 + \frac{iD}{\omega}\big[q_{\parallel}^2 + q_{\perp 1}^2 + q_{\perp 2}^2\big] + \frac{D\,ze{E}_{\parallel}}{\omega\,k_BT}q_{\parallel}},
\end{split}
\end{equation}
with $q_{\perp 1}$ and $q_{\perp 2}$ being the two perpendicular wave vectors.
The power spectrum of the noise is proportional to the bulk concentration $C_0$,
\begin{equation} \label{eq:er17}
\begin{split}
\langle \eta_{\gamma}\left(\myvec{x},t\right)\eta_{\beta}(\myvec{x}',t')\rangle  &=  2DC_{0}\delta_{\gamma\beta}\delta(\myvec{x}-\myvec{x}')\delta(t-t') \\
\langle \widetilde{\eta}_{\gamma}(\myvec{q},\omega)\widetilde{\eta}_{\beta}(\myvec{q}',\omega')\rangle &= 2DC_{0}\delta_{\gamma\beta}(2\pi)^{4}\delta(\myvec{q}+\myvec{q}') \delta(\omega+\omega').
\end{split}
\end{equation}
Using short-hand notation, we derive from Eqs. \ref{eq:er14}-\ref{eq:er17}
\begin{equation} \label{eq:er18}
\begin{split}
\langle| & \widetilde{J}_{\parallel}^{+} - \widetilde{J}_{\parallel}^{-}|^2\rangle \equiv \int\!\!\int\!\!\int \frac{\textrm{d}q_{\perp1}'\textrm{d}q_{\perp2}'\textrm{d}q_{\parallel}'}{(2\pi)^3}\int \frac{\textrm{d}\omega'}{2\pi} \Big[\\
\langle\big[ &\widetilde{J}^{+}_{\parallel}(\myvec{q},\omega)-\widetilde{J}^{-}_{\parallel}(\myvec{q},\omega)\big]
       \big[\widetilde{J}^{+}_{\parallel}(\myvec{q}',\omega')-\widetilde{J}^{-}_{\parallel}(\myvec{q}',\omega')\big]\rangle \Big] = \\
& \frac{8 D C_{0} \Big[\frac{eE_{\parallel}}{kT}\Big]^2 \Big[\frac{\omega^2}{D^2} + q_{\perp}^{4}\Big] \Big[q_{\perp}^{2} + q_{\parallel}^{2}\Big]}
       {\Big(\Big[q_{\perp}^{2} + q_{\parallel}^{2}\Big]^2  \!\! + \Big[\frac{ eE_{\parallel} }{kT}\Big]^2 \! q_{\parallel}^2 \! - \frac{\omega^2}{D^2} \! \Big)^2 
     \!\!  + 4 \frac{\omega^2}{D^2} \Big[q_{\perp}^{2} + q_{\parallel}^{2}\Big]^2}.
\end{split}
\end{equation}

The power spectral density $S\left(\omega\right)$ of the current $I_{\parallel}\left(t\right)$ defined on the domain $0 < t <T$ is given by the limit of $T \to \infty$ of
\begin{equation}
S\left(\omega\right) = \frac{1}{T} \langle | \tilde{I}_{\parallel}\left(\omega\right) |^2 \rangle = \frac{1}{T} \langle \tilde{I}_{\parallel} \left(\omega\right) \tilde{I}_{\parallel}\left(-\omega\right) \rangle,
\end{equation}
which can be written as
\begin{equation}
S\left(\omega\right) = \frac{1}{T} \int_T \!\textrm{d}t \int_T \!\textrm{d}t' \, e^{-i\omega\left(t-t'\right)} \, \langle I_{\parallel}\left(t\right) I_{\parallel}\left(t'\right) \rangle.
\end{equation}
We write $I_{\parallel}\left(t\right)$ as the integral of the current density $J^+_{\parallel}\left(\myvec{x},t\right)-J^-_{\parallel}\left(\myvec{x},t\right)$ at a given position in the direction of $x_{\parallel}$
over the lateral surface area $A$ of the pore,
\begin{equation} \label{eqn:S_of_J}
\begin{split}
& S\left(\omega\right) = \frac{1}{T} \int_T \!\!\textrm{d}t \! \int_T \!\!\textrm{d}t' 
\, e^{-i\omega\left(t-t'\right)} \\
& \! \int\!\!\!\int_A \!\!\textrm{d}x_{\perp1}  \textrm{d}x_{\perp2} \! \int \!\!\textrm{d}x_{\parallel}
  \! \int\!\!\!\int_A \!\!\textrm{d}x_{\perp1}' \textrm{d}x_{\perp2}'\! \int \!\!\textrm{d}x_{\parallel}' \delta(x_{\parallel})\delta(x_{\parallel}')\\
&\langle 
\big[J^+_{\parallel}\left(\myvec{x},t\right) - J^-_{\parallel}\left(\myvec{x},t\right)\big]
\big[J^+_{\parallel}\left(\myvec{x}',t'\right) - J^-_{\parallel}\left(\myvec{x}',t'\right)\big]
\rangle.
\end{split}
\end{equation}
Expressing the delta functions and the current density in Eq. \ref{eqn:S_of_J} in terms of their Fourier transforms, we arrive at
\begin{equation}
\begin{split}
& S\left(\omega\right) = \frac{1}{T} \int_T \!\!\textrm{d}t \! \int_T \!\!\textrm{d}t' 
\, e^{-i\omega\left(t-t'\right)} \\
& \! \int\!\!\!\int_A \!\!\textrm{d}x_{\perp1}  \textrm{d}x_{\perp2} \! \int \!\!\textrm{d}x_{\parallel} 
  \! \int\!\!\!\int_A \!\!\textrm{d}x_{\perp1}' \textrm{d}x_{\perp2}'\! \int \!\!\textrm{d}x_{\parallel}'\\
& \! \int \! \frac{\textrm{d}q_{\perp1} }{2\pi} \! \int \! \frac{\textrm{d}q_{\perp2} }{2\pi} \! \int \! \frac{\textrm{d}q_{\parallel} }{2\pi}
  \, e^{-i(q_{\perp1}x_{\perp1}+q_{\perp2}x_{\perp2}+q_{\parallel}x_{\parallel})} \\
& \! \int \! \frac{\textrm{d}q_{\perp1}' }{2\pi} \! \int \! \frac{\textrm{d}q_{\perp2}' }{2\pi} \! \int \! \frac{\textrm{d}q_{\parallel}' }{2\pi}
  \, e^{-i(q_{\perp1}'x_{\perp1}'+q_{\perp2}'x_{\perp2}'+q_{\parallel}'x_{\parallel}')} \\
& \! \int \! \frac{\textrm{d}\omega'}{2\pi} \!\int\!\frac{\textrm{d}\omega''}{2\pi} \, e^{i\omega' t} e^{i\omega'' t'}
  \! \int\frac{\textrm{d}q_{\parallel}''}{2\pi} e^{iq_{\parallel}'' x_{\parallel}} \!\! \int\frac{\textrm{d}q_{\parallel}'''}{2\pi} e^{iq_{\parallel}''' x_{\parallel}'} \\
&\langle
\big[\widetilde{J}^+_{\parallel}\left(\myvec{q},\omega'\right) - \widetilde{J}^-_{\parallel}\left(\myvec{q},\omega'\right)\big]
\big[\widetilde{J}^+_{\parallel}\left(\myvec{q}',\omega''\right) - \widetilde{J}^-_{\parallel}\left(\myvec{q}',\omega''\right)\big]
\rangle.
\end{split}
\end{equation}
Performing the integrals over $\omega''$, $\myvec{q}'$, $q_{\parallel}''$, $q_{\parallel}'''$, $x_{\parallel}$, and $x_{\parallel}'$, and using the short-hand notation of Eq. \ref{eq:er18} leads to
\begin{equation}
\begin{split}
& S\left(\omega\right) = \frac{1}{T} \int_T \!\!\textrm{d}t \! \int_T \!\!\textrm{d}t' 
\, e^{-i\omega\left(t-t'\right)} \\
& \! \int\!\!\!\int_A \!\!\textrm{d}x_{\perp1}  \textrm{d}x_{\perp2} 
  \! \int\!\!\!\int_A \!\!\textrm{d}x_{\perp1}' \textrm{d}x_{\perp2}'\\
& \! \int \! \frac{\textrm{d}q_{\perp1} }{2\pi} \! \int \! \frac{\textrm{d}q_{\perp2} }{2\pi} \! \int \! \frac{\textrm{d}q_{\parallel} }{2\pi}
  e^{-i(q_{\perp1}(x_{\perp1}-x_{\perp1}')+q_{\perp2}(x_{\perp2}-x_{\perp2}'))} \\
& \! \int \! \frac{\textrm{d}\omega'}{2\pi} e^{i\omega' (t-t')} 
 \langle |\big[\tilde{J}^+_{\parallel}\left( \myvec{q}, \omega'\right) - \tilde{J}^-_{\parallel}\left( \myvec{q}, \omega'\right)\big] |^2 \rangle.
\end{split}
\end{equation}
We rearrange the exponential functions
and perform the integrals over $t$, $t'$ and $\omega'$, yielding
\begin{equation} \label{eqn:psd-derivation}
\begin{split}
S\left(\omega\right) = \int \! \frac{\textrm{d}q_{\perp1} }{2\pi} \! & \int \! \frac{\textrm{d}q_{\perp2} }{2\pi} \! \int \! \frac{\textrm{d}q_{\parallel} }{2\pi} \widetilde{A}^2(q_{\perp}) \\
& \langle |\big[\tilde{J}^+_{\parallel}\left( \myvec{q}, \omega\right) - \tilde{J}^-_{\parallel}\left( \myvec{q}, \omega\right)\big] |^2 \rangle,
\end{split}
\end{equation}
with the squared Fourier-transformed area function being given by
\begin{equation} \label{eqn:area-derivation}
\begin{split}
\widetilde{A}^2(q_{\perp}) & =
  \! \int\!\!\!\int_A \!\!\textrm{d}x_{\perp1}  \textrm{d}x_{\perp2} 
  \! \int\!\!\!\int_A \!\!\textrm{d}x_{\perp1}' \textrm{d}x_{\perp2}' \\
& \quad\quad\quad\quad e^{-i(q_{\perp1}(x_{\perp1}-x_{\perp1}')+q_{\perp2}(x_{\perp2}-x_{\perp2}'))} \\
& = \int_{0}^{2\pi}\!\!\!\textrm{d}\theta \! \int_{0}^{2\pi}\!\!\! \textrm{d}\theta' \! \int_{0}^{R}\!\!\! \textrm{d}x_{\perp} \! \int_{0}^{R} \!\!\! \textrm{d}x_{\perp}' \, x_{\perp} \, x_{\perp}' \, e^{-iq_{\perp}\left(x_{\perp}-x_{\perp}'\right)},
\end{split}
\end{equation}
with $x_{\perp} = \sqrt{x_{\perp1}^2 + x_{\perp2}^2}$ and $\theta = \arctan(x_{\perp2}/x_{\perp1})$ being the cylindrical coordinates and $q_{\perp} = \sqrt{q_{\perp1}^2+q_{\perp2}^2}$.
Combining Eqs. \ref{eq:er18} and \ref{eqn:psd-derivation}, and
performing the integrals in Eq. \ref{eqn:area-derivation}, we arrive at Eqs. \ref{eqn:psd}-\ref{eqn:current}.

\begin{figure}
\includegraphics[width=0.5\columnwidth]{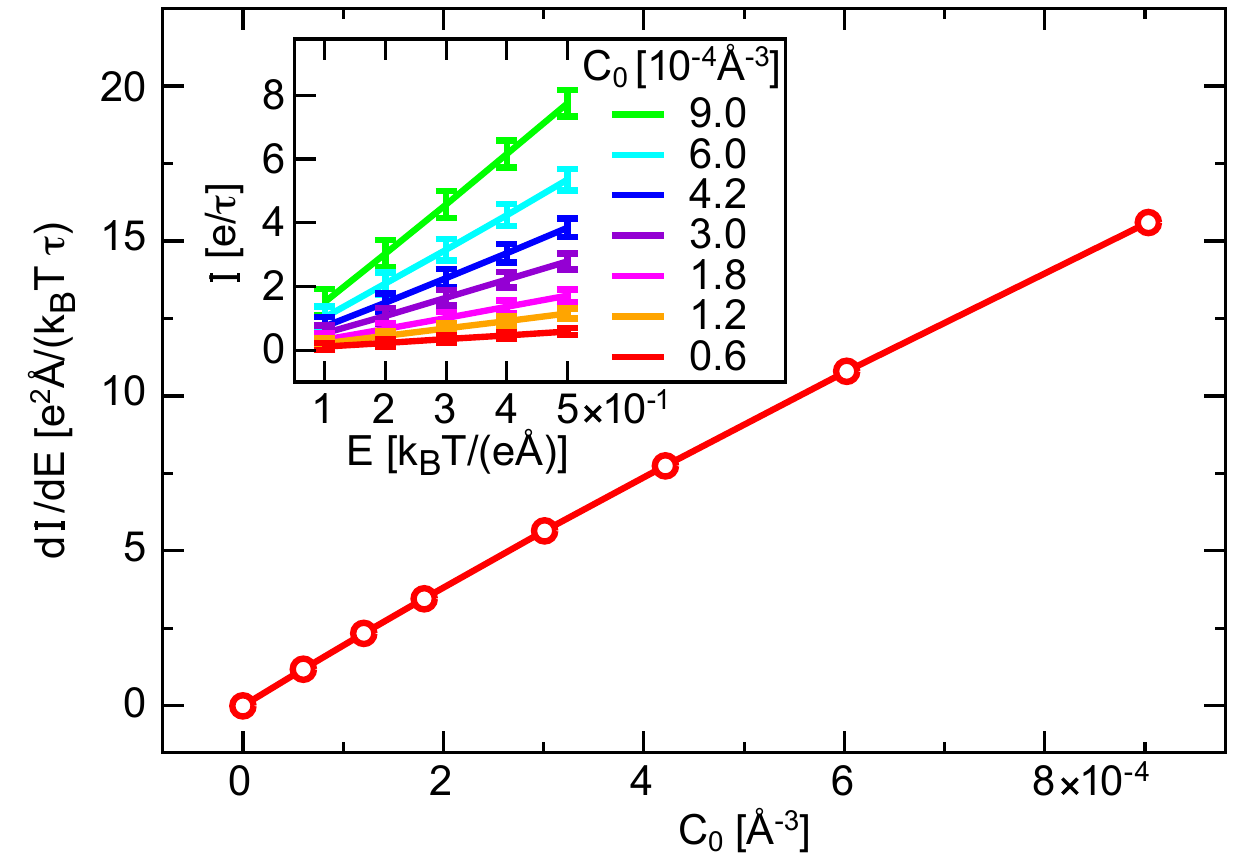}
\caption{
The electrical conductivity of a bulk system of ions ($100\times100\times100$~\AA) as a function of the ionic concentration $C_0$.
Inset: The current as a function of the electric field.
}
\label{fig:conductivity}
\end{figure}

\textbf{Time scale.} --
We calibrate the time scale $\tau$ from the conductivity of a bulk system of $100^3$~\AA$^3$, calculated from a linear fit of the current as a function of the electric field $E$ (Fig. \ref{fig:conductivity}).
Experimentally, the conductivity is given by
\begin{equation} \label{eqn:conductivity}
\frac{1}{C_0}\frac{\textrm{d}I}{\textrm{d}E} = A\left(D^+ + D^-\right)\frac{e^2}{k_BT},
\end{equation}
with $A$ the lateral surface area and $D^{+} = D^{-} = 0.2$~\AA$^2$/ps being the diffusion coefficient of K$^+$ and Cl$^-$.
We calculate the time scale $\tau$ from a linear fit to the curve in Fig. \ref{fig:conductivity} up to $C_0 = 3\cdot 10^{-4}$ and comparison with Eq. \ref{eqn:conductivity}, giving $\tau = 5$~ps.

\begin{figure}
\includegraphics[width=0.5\columnwidth]{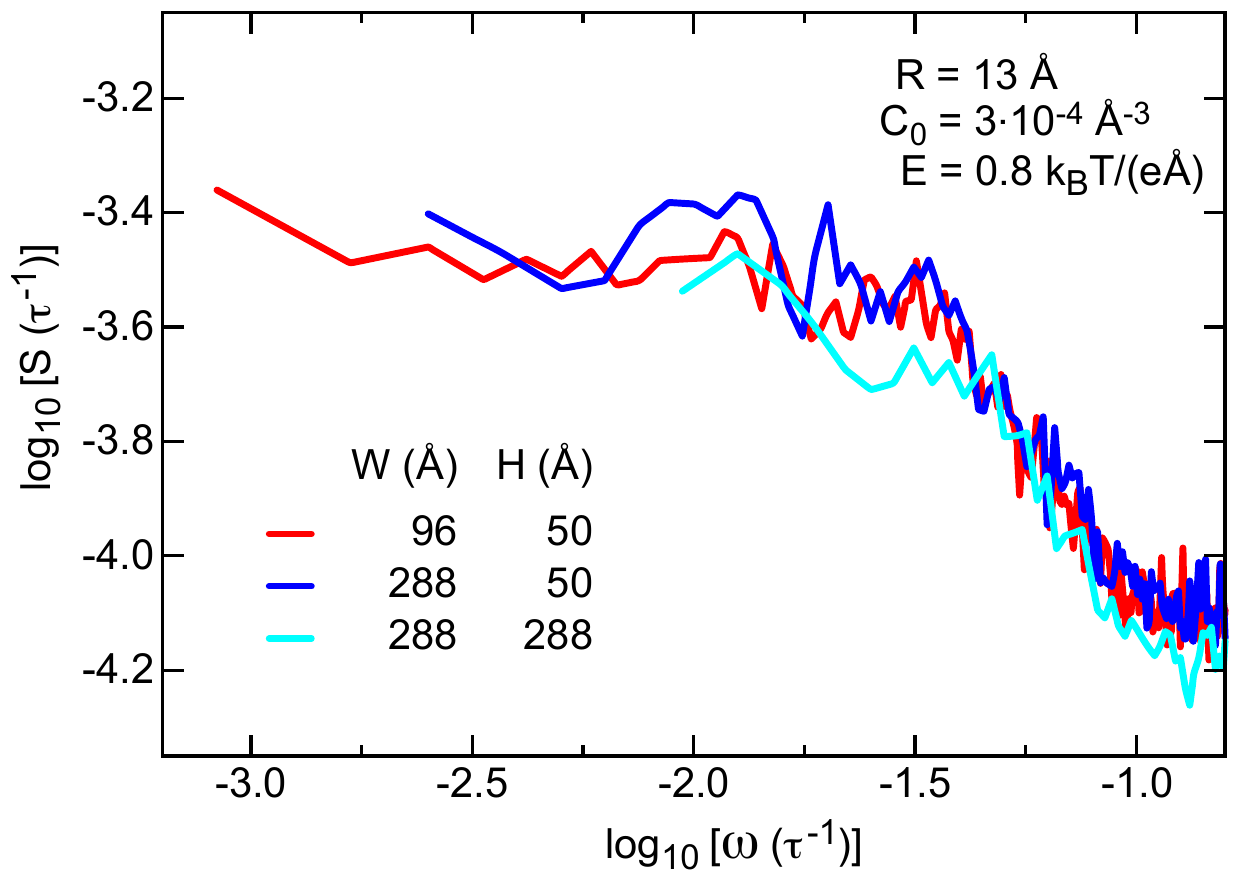}
\caption{
The power spectra of pores with radius $R = 13$~\AA and length $L = 48$~\AA, each connected to two reservoirs of size $W\times W\times H$.
The total simulation box is $W \times W \times L+2H$, with periodic boundary conditions in all directions.
The simulations are performed with $V_{\textit{ion-ion}} = 0$.
}
\label{fig:size-dependence}
\end{figure}

\textbf{Finite-size effects.} --
To check the effect of the simulation box size, we vary the reservoir size between $50$~\AA~and $288$~\AA~in $x_{\parallel}$-direction and between $96$~\AA~and $288$~\AA~in $x_{\perp}$-direction.
The power spectra are independent of the reservoir size (Fig. \ref{fig:size-dependence}), from which we conclude that the periodic boundary conditions do not affect the ion fluctuations.


\input{noise_v04b.bbl}

\end{document}

%% file: noise_v04b.bbl
%

%% file: noise_v04c.bbl
\begin{thebibliography}{23}%
\makeatletter
\providecommand \@ifxundefined [1]{%
 \@ifx{#1\undefined}
}%
\providecommand \@ifnum [1]{%
 \ifnum #1\expandafter \@firstoftwo
 \else \expandafter \@secondoftwo
 \fi
}%
\providecommand \@ifx [1]{%
 \ifx #1\expandafter \@firstoftwo
 \else \expandafter \@secondoftwo
 \fi
}%
\providecommand \natexlab [1]{#1}%
\providecommand \enquote  [1]{``#1''}%
\providecommand \bibnamefont  [1]{#1}%
\providecommand \bibfnamefont [1]{#1}%
\providecommand \citenamefont [1]{#1}%
\providecommand \href@noop [0]{\@secondoftwo}%
\providecommand \href [0]{\begingroup \@sanitize@url \@href}%
\providecommand \@href[1]{\@@startlink{#1}\@@href}%
\providecommand \@@href[1]{\endgroup#1\@@endlink}%
\providecommand \@sanitize@url [0]{\catcode `\\12\catcode `\$12\catcode
  `\&12\catcode `\#12\catcode `\^12\catcode `\_12\catcode `\%12\relax}%
\providecommand \@@startlink[1]{}%
\providecommand \@@endlink[0]{}%
\providecommand \url  [0]{\begingroup\@sanitize@url \@url }%
\providecommand \@url [1]{\endgroup\@href {#1}{\urlprefix }}%
\providecommand \urlprefix  [0]{URL }%
\providecommand \Eprint [0]{\href }%
\providecommand \doibase [0]{http://dx.doi.org/}%
\providecommand \selectlanguage [0]{\@gobble}%
\providecommand \bibinfo  [0]{\@secondoftwo}%
\providecommand \bibfield  [0]{\@secondoftwo}%
\providecommand \translation [1]{[#1]}%
\providecommand \BibitemOpen [0]{}%
\providecommand \bibitemStop [0]{}%
\providecommand \bibitemNoStop [0]{.\EOS\space}%
\providecommand \EOS [0]{\spacefactor3000\relax}%
\providecommand \BibitemShut  [1]{\csname bibitem#1\endcsname}%
\let\auto@bib@innerbib\@empty
\bibitem [{\citenamefont {Dekker}(2007)}]{2007_Dekker}%
  \BibitemOpen
  \bibfield  {author} {\bibinfo {author} {\bibfnamefont {Cees}\ \bibnamefont
  {Dekker}},\ }\bibfield  {title} {\enquote {\bibinfo {title} {Solid-state
  nanopores},}\ }\href@noop {} {\bibfield  {journal} {\bibinfo  {journal} {Nat.
  Nanotechol.}\ }\textbf {\bibinfo {volume} {2}},\ \bibinfo {pages} {209--215}
  (\bibinfo {year} {2007})}\BibitemShut {NoStop}%
\bibitem [{\citenamefont {Kasianowicz}\ \emph {et~al.}(1996)\citenamefont
  {Kasianowicz} \emph {et~al.}}]{1996_Kasianowicz}%
  \BibitemOpen
  \bibfield  {author} {\bibinfo {author} {\bibfnamefont {J.}~\bibnamefont
  {Kasianowicz}} \emph {et~al.},\ }\bibfield  {title} {\enquote {\bibinfo
  {title} {Characterization of individual polynucleotide molecules using a
  membrane channel},}\ }\href@noop {} {\bibfield  {journal} {\bibinfo
  {journal} {Proc. Nat. Acad. Sci. USA}\ }\textbf {\bibinfo {volume} {93}},\
  \bibinfo {pages} {13770--13773} (\bibinfo {year} {1996})}\BibitemShut
  {NoStop}%
\bibitem [{\citenamefont {Deamer}\ and\ \citenamefont
  {Akeson}(2000)}]{2000_Deamer}%
  \BibitemOpen
  \bibfield  {author} {\bibinfo {author} {\bibfnamefont {D.~W.}\ \bibnamefont
  {Deamer}}\ and\ \bibinfo {author} {\bibfnamefont {M.}~\bibnamefont
  {Akeson}},\ }\bibfield  {title} {\enquote {\bibinfo {title} {Nanopores and
  nucleic acids: prospects for ultrarapid sequencing},}\ }\href@noop {}
  {\bibfield  {journal} {\bibinfo  {journal} {Trends Biotechnol.}\ }\textbf
  {\bibinfo {volume} {18}},\ \bibinfo {pages} {131--180} (\bibinfo {year}
  {2000})}\BibitemShut {NoStop}%
\bibitem [{\citenamefont {Meller}\ \emph {et~al.}(2000)\citenamefont {Meller},
  \citenamefont {Nivon}, \citenamefont {Brandin}, \citenamefont {Golovchenko},\
  and\ \citenamefont {Branton}}]{2000_Meller}%
  \BibitemOpen
  \bibfield  {author} {\bibinfo {author} {\bibfnamefont {Amit}\ \bibnamefont
  {Meller}}, \bibinfo {author} {\bibfnamefont {Lucas}\ \bibnamefont {Nivon}},
  \bibinfo {author} {\bibfnamefont {Eric}\ \bibnamefont {Brandin}}, \bibinfo
  {author} {\bibfnamefont {Jene}\ \bibnamefont {Golovchenko}}, \ and\ \bibinfo
  {author} {\bibfnamefont {Daniel}\ \bibnamefont {Branton}},\ }\bibfield
  {title} {\enquote {\bibinfo {title} {Rapid nanopore discrimination between
  single polynucleotide molecules},}\ }\href@noop {} {\bibfield  {journal}
  {\bibinfo  {journal} {Proc. Nat. Acad. Sci. USA}\ }\textbf {\bibinfo {volume}
  {97}},\ \bibinfo {pages} {1079--1084} (\bibinfo {year} {2000})}\BibitemShut
  {NoStop}%
\bibitem [{\citenamefont {Weissman}(1988)}]{1988_Weissman_RevModPhys}%
  \BibitemOpen
  \bibfield  {author} {\bibinfo {author} {\bibfnamefont {M.~B.}\ \bibnamefont
  {Weissman}},\ }\bibfield  {title} {\enquote {\bibinfo {title} {1/f noise and
  other slow, non-exponential kinetics in condensed matter},}\ }\href@noop {}
  {\bibfield  {journal} {\bibinfo  {journal} {Rev. Mod. Phys.}\ }\textbf
  {\bibinfo {volume} {60}},\ \bibinfo {pages} {537--571} (\bibinfo {year}
  {1988})}\BibitemShut {NoStop}%
\bibitem [{\citenamefont {Singh}\ \emph {et~al.}(2011)\citenamefont {Singh},
  \citenamefont {Chan}, \citenamefont {Kang},\ and\ \citenamefont
  {Lemay}}]{2011_Singh-Lemay}%
  \BibitemOpen
  \bibfield  {author} {\bibinfo {author} {\bibfnamefont {Pradyumna~S.}\
  \bibnamefont {Singh}}, \bibinfo {author} {\bibfnamefont {{Hui-Shan}~M.}\
  \bibnamefont {Chan}}, \bibinfo {author} {\bibfnamefont {Shuo}\ \bibnamefont
  {Kang}}, \ and\ \bibinfo {author} {\bibfnamefont {Serge~G.}\ \bibnamefont
  {Lemay}},\ }\bibfield  {title} {\enquote {\bibinfo {title} {Stochastic
  amperometric fluctuations as a probe for dynamic adsorption in nanofluidic
  electrochemical systems},}\ }\href@noop {} {\bibfield  {journal} {\bibinfo
  {journal} {J. Am. Chem. Soc.}\ }\textbf {\bibinfo {volume} {133}},\ \bibinfo
  {pages} {18289--18295} (\bibinfo {year} {2011})}\BibitemShut {NoStop}%
\bibitem [{\citenamefont {Heerema}\ \emph {et~al.}(2015)\citenamefont
  {Heerema}, \citenamefont {Schneider}, \citenamefont {Rozemuller},
  \citenamefont {Vicarelli}, \citenamefont {Zandbergen},\ and\ \citenamefont
  {Dekker}}]{2015_Heerema}%
  \BibitemOpen
  \bibfield  {author} {\bibinfo {author} {\bibfnamefont {SJ}~\bibnamefont
  {Heerema}}, \bibinfo {author} {\bibfnamefont {GF}~\bibnamefont {Schneider}},
  \bibinfo {author} {\bibfnamefont {M}~\bibnamefont {Rozemuller}}, \bibinfo
  {author} {\bibfnamefont {L}~\bibnamefont {Vicarelli}}, \bibinfo {author}
  {\bibfnamefont {HW}~\bibnamefont {Zandbergen}}, \ and\ \bibinfo {author}
  {\bibfnamefont {C}~\bibnamefont {Dekker}},\ }\bibfield  {title} {\enquote
  {\bibinfo {title} {1/f noise in graphene nanopores},}\ }\href@noop {}
  {\bibfield  {journal} {\bibinfo  {journal} {Nanotechnology}\ }\textbf
  {\bibinfo {volume} {26}},\ \bibinfo {pages} {074001} (\bibinfo {year}
  {2015})}\BibitemShut {NoStop}%
\bibitem [{\citenamefont {Smeets}\ \emph {et~al.}(2008)\citenamefont {Smeets},
  \citenamefont {Keyser}, \citenamefont {Dekker},\ and\ \citenamefont
  {Dekker}}]{2008_Smeets_PNAS}%
  \BibitemOpen
  \bibfield  {author} {\bibinfo {author} {\bibfnamefont {Ralph~MM}\
  \bibnamefont {Smeets}}, \bibinfo {author} {\bibfnamefont {Ulrich~F}\
  \bibnamefont {Keyser}}, \bibinfo {author} {\bibfnamefont {Nynke~H}\
  \bibnamefont {Dekker}}, \ and\ \bibinfo {author} {\bibfnamefont {Cees}\
  \bibnamefont {Dekker}},\ }\bibfield  {title} {\enquote {\bibinfo {title}
  {Noise in solid-state nanopores},}\ }\href@noop {} {\bibfield  {journal}
  {\bibinfo  {journal} {Proc. Nat. Acad. Sci. USA}\ }\textbf {\bibinfo {volume}
  {105}},\ \bibinfo {pages} {417--421} (\bibinfo {year} {2008})}\BibitemShut
  {NoStop}%
\bibitem [{\citenamefont {Tabard-Cossa}\ \emph {et~al.}(2007)\citenamefont
  {Tabard-Cossa}, \citenamefont {Trivedi}, \citenamefont {Wiggin},
  \citenamefont {Jetha},\ and\ \citenamefont {Marziali}}]{2007_Tabard-Cossa}%
  \BibitemOpen
  \bibfield  {author} {\bibinfo {author} {\bibfnamefont {Vincent}\ \bibnamefont
  {Tabard-Cossa}}, \bibinfo {author} {\bibfnamefont {D.}~\bibnamefont
  {Trivedi}}, \bibinfo {author} {\bibfnamefont {M.}~\bibnamefont {Wiggin}},
  \bibinfo {author} {\bibfnamefont {N.~N.}\ \bibnamefont {Jetha}}, \ and\
  \bibinfo {author} {\bibfnamefont {A.}~\bibnamefont {Marziali}},\ }\bibfield
  {title} {\enquote {\bibinfo {title} {Noise analysis and reduction in
  solid-state nanopores},}\ }\href@noop {} {\bibfield  {journal} {\bibinfo
  {journal} {Nanotechnology}\ }\textbf {\bibinfo {volume} {18}},\ \bibinfo
  {pages} {305505} (\bibinfo {year} {2007})}\BibitemShut {NoStop}%
\bibitem [{\citenamefont {Bezrukov}\ and\ \citenamefont
  {Winterhalter}(2000)}]{2000_Bezrukov_PRL}%
  \BibitemOpen
  \bibfield  {author} {\bibinfo {author} {\bibfnamefont {Sergey~M}\
  \bibnamefont {Bezrukov}}\ and\ \bibinfo {author} {\bibfnamefont {Mathias}\
  \bibnamefont {Winterhalter}},\ }\bibfield  {title} {\enquote {\bibinfo
  {title} {Examining noise sources at the single-molecule level: 1/f noise of
  an open maltoporin channel},}\ }\href@noop {} {\bibfield  {journal} {\bibinfo
   {journal} {Phys. Rev. Lett.}\ }\textbf {\bibinfo {volume} {85}},\ \bibinfo
  {pages} {202} (\bibinfo {year} {2000})}\BibitemShut {NoStop}%
\bibitem [{\citenamefont {Wohnsland}\ and\ \citenamefont
  {Benz}(1997)}]{1997_Wohnsland}%
  \BibitemOpen
  \bibfield  {author} {\bibinfo {author} {\bibfnamefont {F}~\bibnamefont
  {Wohnsland}}\ and\ \bibinfo {author} {\bibfnamefont {R}~\bibnamefont
  {Benz}},\ }\bibfield  {title} {\enquote {\bibinfo {title} {1/f-noise of open
  bacterial porin channels},}\ }\href@noop {} {\bibfield  {journal} {\bibinfo
  {journal} {J. Membr. Biol.}\ }\textbf {\bibinfo {volume} {158}},\ \bibinfo
  {pages} {77--85} (\bibinfo {year} {1997})}\BibitemShut {NoStop}%
\bibitem [{\citenamefont {Siwy}\ and\ \citenamefont
  {Fuli{\'n}ski}(2002)}]{2002_Siwy-Fulinski_PRL}%
  \BibitemOpen
  \bibfield  {author} {\bibinfo {author} {\bibfnamefont {Z}~\bibnamefont
  {Siwy}}\ and\ \bibinfo {author} {\bibfnamefont {A}~\bibnamefont
  {Fuli{\'n}ski}},\ }\bibfield  {title} {\enquote {\bibinfo {title} {Origin of
  {1/f$^\alpha$} noise in membrane channel currents},}\ }\href@noop {}
  {\bibfield  {journal} {\bibinfo  {journal} {Phys. Rev. Lett.}\ }\textbf
  {\bibinfo {volume} {89}},\ \bibinfo {pages} {158101} (\bibinfo {year}
  {2002})}\BibitemShut {NoStop}%
\bibitem [{\citenamefont {Hoogerheide}\ \emph {et~al.}(2009)\citenamefont
  {Hoogerheide}, \citenamefont {Garaj},\ and\ \citenamefont
  {Golovchenko}}]{2009_Hoogerheide}%
  \BibitemOpen
  \bibfield  {author} {\bibinfo {author} {\bibfnamefont {David~P}\ \bibnamefont
  {Hoogerheide}}, \bibinfo {author} {\bibfnamefont {Slaven}\ \bibnamefont
  {Garaj}}, \ and\ \bibinfo {author} {\bibfnamefont {Jene~A}\ \bibnamefont
  {Golovchenko}},\ }\bibfield  {title} {\enquote {\bibinfo {title} {Probing
  surface charge fluctuations with solid-state nanopores},}\ }\href@noop {}
  {\bibfield  {journal} {\bibinfo  {journal} {Phys. Rev. Lett.}\ }\textbf
  {\bibinfo {volume} {102}},\ \bibinfo {pages} {256804} (\bibinfo {year}
  {2009})}\BibitemShut {NoStop}%
\bibitem [{\citenamefont {Powell}\ \emph {et~al.}(2009)\citenamefont {Powell},
  \citenamefont {Vlassiouk}, \citenamefont {Martens},\ and\ \citenamefont
  {Siwy}}]{2009_Powell-Siwy_PRL}%
  \BibitemOpen
  \bibfield  {author} {\bibinfo {author} {\bibfnamefont {Matthew~R}\
  \bibnamefont {Powell}}, \bibinfo {author} {\bibfnamefont {Ivan}\ \bibnamefont
  {Vlassiouk}}, \bibinfo {author} {\bibfnamefont {Craig}\ \bibnamefont
  {Martens}}, \ and\ \bibinfo {author} {\bibfnamefont {Zuzanna~S}\ \bibnamefont
  {Siwy}},\ }\bibfield  {title} {\enquote {\bibinfo {title} {Nonequilibrium 1/f
  noise in rectifying nanopores},}\ }\href@noop {} {\bibfield  {journal}
  {\bibinfo  {journal} {Phys. Rev. Lett.}\ }\textbf {\bibinfo {volume} {103}},\
  \bibinfo {pages} {248104} (\bibinfo {year} {2009})}\BibitemShut {NoStop}%
\bibitem [{\citenamefont {Hooge}(1970)}]{1970_Hooge}%
  \BibitemOpen
  \bibfield  {author} {\bibinfo {author} {\bibfnamefont {F.~N.}\ \bibnamefont
  {Hooge}},\ }\bibfield  {title} {\enquote {\bibinfo {title} {1/f noise in the
  conductance of ions in aqueous solutions},}\ }\href@noop {} {\bibfield
  {journal} {\bibinfo  {journal} {Physics Letters A}\ }\textbf {\bibinfo
  {volume} {33}},\ \bibinfo {pages} {169--170} (\bibinfo {year}
  {1970})}\BibitemShut {NoStop}%
\bibitem [{\citenamefont {Hagen}\ \emph {et~al.}(1997)\citenamefont {Hagen},
  \citenamefont {Pagonabarraga}, \citenamefont {Lowe},\ and\ \citenamefont
  {Frenkel}}]{1997_Hagen}%
  \BibitemOpen
  \bibfield  {author} {\bibinfo {author} {\bibfnamefont {M.~H.~J.}\
  \bibnamefont {Hagen}}, \bibinfo {author} {\bibfnamefont {I.}~\bibnamefont
  {Pagonabarraga}}, \bibinfo {author} {\bibfnamefont {C.~P.}\ \bibnamefont
  {Lowe}}, \ and\ \bibinfo {author} {\bibfnamefont {D.}~\bibnamefont
  {Frenkel}},\ }\bibfield  {title} {\enquote {\bibinfo {title} {Algebraic decay
  of velocity fluctuations in a confined fluid},}\ }\href@noop {} {\bibfield
  {journal} {\bibinfo  {journal} {Phys. Rev. Lett.}\ }\textbf {\bibinfo
  {volume} {78}},\ \bibinfo {pages} {3785} (\bibinfo {year}
  {1997})}\BibitemShut {NoStop}%
\bibitem [{\citenamefont {Jeon}\ and\ \citenamefont
  {Metzler}(2010)}]{2010_Jeon-Metzler}%
  \BibitemOpen
  \bibfield  {author} {\bibinfo {author} {\bibfnamefont {{Jae-Hyung}}\
  \bibnamefont {Jeon}}\ and\ \bibinfo {author} {\bibfnamefont {Ralf}\
  \bibnamefont {Metzler}},\ }\bibfield  {title} {\enquote {\bibinfo {title}
  {Fractional brownian motion and motion governed by the fractional langevin
  equation in confined geometries},}\ }\href@noop {} {\bibfield  {journal}
  {\bibinfo  {journal} {Phys. Rev. E.}\ }\textbf {\bibinfo {volume} {81}},\
  \bibinfo {pages} {021103} (\bibinfo {year} {2010})}\BibitemShut {NoStop}%
\bibitem [{\citenamefont {Tasserit}\ \emph {et~al.}(2010)\citenamefont
  {Tasserit}, \citenamefont {Koutsioubas}, \citenamefont {Lairez},
  \citenamefont {Zalczer},\ and\ \citenamefont {Clochard}}]{2010_Tasserit}%
  \BibitemOpen
  \bibfield  {author} {\bibinfo {author} {\bibfnamefont {C.}~\bibnamefont
  {Tasserit}}, \bibinfo {author} {\bibfnamefont {A.}~\bibnamefont
  {Koutsioubas}}, \bibinfo {author} {\bibfnamefont {D.}~\bibnamefont {Lairez}},
  \bibinfo {author} {\bibfnamefont {G.}~\bibnamefont {Zalczer}}, \ and\
  \bibinfo {author} {\bibfnamefont {{M.-C.}}\ \bibnamefont {Clochard}},\
  }\bibfield  {title} {\enquote {\bibinfo {title} {Pink noise of ionic
  conductance through single artificial nanopores revisited},}\ }\href@noop {}
  {\bibfield  {journal} {\bibinfo  {journal} {Phys. Rev. Lett.}\ }\textbf
  {\bibinfo {volume} {105}},\ \bibinfo {pages} {260602} (\bibinfo {year}
  {2010})}\BibitemShut {NoStop}%
\bibitem [{\citenamefont {Kosi{\'n}ska}\ and\ \citenamefont
  {Fuli{\'n}ski}(2008)}]{2008_Fulinski}%
  \BibitemOpen
  \bibfield  {author} {\bibinfo {author} {\bibfnamefont {I~D}\ \bibnamefont
  {Kosi{\'n}ska}}\ and\ \bibinfo {author} {\bibfnamefont {A}~\bibnamefont
  {Fuli{\'n}ski}},\ }\bibfield  {title} {\enquote {\bibinfo {title} {Brownian
  dynamics simulations of flicker noise in nanochannels currents},}\
  }\href@noop {} {\bibfield  {journal} {\bibinfo  {journal} {Europhys. Lett.}\
  }\textbf {\bibinfo {volume} {81}},\ \bibinfo {pages} {50006} (\bibinfo {year}
  {2008})}\BibitemShut {NoStop}%
\bibitem [{sup()}]{supp}%
  \BibitemOpen
  \href@noop {} {}\bibinfo {note} {Appendix}\BibitemShut {NoStop}%
\bibitem [{\citenamefont {Limbach}\ \emph {et~al.}(2006)\citenamefont
  {Limbach}, \citenamefont {Arnold}, \citenamefont {Mann},\ and\ \citenamefont
  {Holm}}]{2006_Limbach}%
  \BibitemOpen
  \bibfield  {author} {\bibinfo {author} {\bibfnamefont {H-J}\ \bibnamefont
  {Limbach}}, \bibinfo {author} {\bibfnamefont {A.}~\bibnamefont {Arnold}},
  \bibinfo {author} {\bibfnamefont {B.A.}\ \bibnamefont {Mann}}, \ and\
  \bibinfo {author} {\bibfnamefont {C.}~\bibnamefont {Holm}},\ }\bibfield
  {title} {\enquote {\bibinfo {title} {Espresso -- an extensible simulation
  package for research on soft matter systems},}\ }\href@noop {} {\bibfield
  {journal} {\bibinfo  {journal} {Computer Physics Communications}\ }\textbf
  {\bibinfo {volume} {174}},\ \bibinfo {pages} {24} (\bibinfo {year}
  {2006})}\BibitemShut {NoStop}%
\bibitem [{\citenamefont {Villamaina}\ and\ \citenamefont
  {Trizac}(2014)}]{2014_Villamaina-Trizac}%
  \BibitemOpen
  \bibfield  {author} {\bibinfo {author} {\bibfnamefont {Dario}\ \bibnamefont
  {Villamaina}}\ and\ \bibinfo {author} {\bibfnamefont {Emmanuel}\ \bibnamefont
  {Trizac}},\ }\bibfield  {title} {\enquote {\bibinfo {title} {Thinking outside
  the box: fluctuations and finite size effects},}\ }\href@noop {} {\bibfield
  {journal} {\bibinfo  {journal} {Eur. J. Phys.}\ }\textbf {\bibinfo {volume}
  {35}},\ \bibinfo {pages} {035011} (\bibinfo {year} {2014})}\BibitemShut
  {NoStop}%
\bibitem [{\citenamefont {Welch}(1967)}]{1967_Welch}%
  \BibitemOpen
  \bibfield  {author} {\bibinfo {author} {\bibfnamefont {Peter~D.}\
  \bibnamefont {Welch}},\ }\bibfield  {title} {\enquote {\bibinfo {title} {The
  use of fast fourier transformfor the estimation of power spectra: A method
  based on time averaging over short, modified periodograms},}\ }\href@noop {}
  {\bibfield  {journal} {\bibinfo  {journal} {IEEE Trans. Audio
  Electroacoust.}\ }\textbf {\bibinfo {volume} {AU-15}},\ \bibinfo {pages}
  {70--73} (\bibinfo {year} {1967})}\BibitemShut {NoStop}%
\end{thebibliography}
